\documentclass[italian,english]{article}
\usepackage[latin9]{inputenc}
\usepackage{color}
\usepackage{graphicx}
\usepackage{amssymb}

\newcommand{\lyxaddress}[1]{
\par {\raggedright #1
\vspace{1.4em}
\noindent\par}
}

\usepackage{babel}

\begin{document}

\title{\textbf{Massive relic gravitational waves from f(R) theories of gravity:
production and potential detection}}

\author{\textbf{Christian Corda }}

\maketitle

\lyxaddress{\begin{center}
Associazione Galileo Galilei, Via Pier Cironi 16 - 59100 PRATO, Italy 
\par\end{center}}

\begin{center}
\textit{E-mail address:} \textcolor{blue}{christian.corda@ego-gw.it} 
\par\end{center}
\begin{abstract}
The production of a stochastic background of relic gravitational waves
is well known in various works in the literature, where, by using
the so called adiabatically-amplified zero-point fluctuations process,
it has been shown how the standard inflationary scenario for the early
universe can in principle provide a distinctive spectrum of relic
gravitational waves. In this paper, it is shown that, in general,
f(R) theories of gravity produce a third massive polarization of gravitational
waves and the primordial production of this polarization is analysed
adapting the adiabatically-amplified zero-point fluctuations process
at this case. In this way, previous results, where only particular
cases of f(R) theories have been analysed, will be generalized.

The presence of the mass could also have important applications in
cosmology, because the fact that gravitational waves can have mass
could give a contribution to the dark matter of the Universe. 

An upper bound for these relic gravitational waves, which arises from
the WMAP constrains, is also released.

At the end of the paper, the potential detection of such massive gravitational
waves using interferometers like Virgo and LIGO is discussed.
\end{abstract}

\section{Introduction}

Recently, the data analysis of interferometric gravitational waves
(GWs) detectors has been started (for the current status of GWs interferometers
see \cite{key-1,key-2,key-3,key-4,key-5,key-6,key-7,key-8}) and the
scientific community aims in a first direct detection of GWs in next
years. 

Detectors for GWs will be important for a better knowledge of the
Universe and also to confirm or to rule out the physical consistency
of General Relativity or of any other theory of gravitation \cite{key-9,key-10,key-11,key-12,key-13,key-14}.
In fact, in the context of Extended Theories of Gravity, some differences
between General Relativity and the others theories can be pointed
out starting by the linearized theory of gravity \cite{key-9,key-10,key-12,key-14}. 

In this picture, detectors for GWs are in principle sensitive also
to a hypothetical \textit{scalar} component of gravitational radiation,
that appears in extended theories of gravity like scalar-tensor gravity
and high order theories \cite{key-12,key-15,key-16,key-17,key-18,key-19,key-20,key-21,key-22},
Brans-Dicke theory \cite{key-23} and string theory \cite{key-24}.

A possible target of these experiments is the so called stochastic
background of gravitational waves \cite{key-25,key-26,key-27,key-28,key-29,key-30}. 

The production of the primordial part of this stochastic background
(relic GWs) is well known in the literature starting by the works
of \cite{key-25,key-26} and \cite{key-27,key-28}, that, using the
so called adiabatically-amplified zero-point fluctuations process,
have shown in two different ways how the standard inflationary scenario
for the early universe can in principle provide a distinctive spectrum
of relic gravitational waves. In \cite{key-29,key-30} the primordial
production has been analysed for the scalar component admitted from
scalar-tensor gravity. 

In this paper, it is shown that, in general, f(R) theories of gravity
produce a third massive polarization of gravitational waves and the
primordial production of this polarization is analysed adapting the
adiabatically-amplified zero-point fluctuations process at this case.
In a recent paper \cite{key-21}, such a process has been applied
to the same class of theories, i.e. the $f(R)$ ones, which will be
discussed in the present work. But, in \cite{key-21} a different
point of view has been considered. In that case, by using a conform
analysis, the authors discussed such a process in respect to the two
standard polarizations which arises from standard General Relativity.
In the present paper the analysis is focused to the third massive
polarization. In this way, we generalize previous results where only
particular cases of f(R) theories have been analysed (see \cite{key-40}
for example).

Regarding f(R) theories, even if such theories have to be dismissed
either because they contradict the current cosmology and solar system
tests or generate \char`\"{}ghosts\char`\"{} in what would be a quantum
version of the finite range gravity\cite{key-41}, a number of authors
manage to avoid both of such problems \cite{key-41}. 

Concerning the presence of the graviton mass, it was considered by
many authors since \foreignlanguage{italian}{Fierz} and Pauli (1939)
\cite{key-42,key-43}. Important contributions in this sense are the
ones in \cite{key-44,key-45,key-46}. In these papers, it is emphasized
that the presence of the mass modifies General Relativity on various
scales, which are defined by the Compton wavelength of gravitons.
This fact leads to dispersion of waves and Yukawa-like potentials
in the linearized theory. On the other hand, it also leads to other
modifications in General Relativity solutions in high-order (i.e.
non-linear) regimes. The fact that gravitational waves can have mass
could also have important applications in cosmology because such masses
could give a contribution to the dark matter of the Universe. 

An upper bound for these relic gravitational waves, which arises from
the WMAP constrains, is also released. About this point, a different
interesting treatment of the additional polarizations and impact on
Cosmic Microwave Background has been recently discussed in the good
paper \cite{key-47}.

At the end of the paper the potential detection of such massive gravitational
waves using interferometers like Virgo and LIGO is discussed.

\section{f(R) theories of gravity}

Let us consider the action\begin{equation}
S=\int d^{4}x\sqrt{-g}f(R)+\mathcal{L}_{m},\label{eq: high order 1}\end{equation}

where $R$ is the Ricci curvature scalar.

Equation (\ref{eq: high order 1}) represents the action of the so
called $f(R)$ theories of gravity \cite{key-9,key-10,key-11,key-19,key-21,key-29,key-33,key-34,key-35}
in respect to the well known canonical one of General Relativity (the
Einstein - Hilbert action \cite{key-31,key-32}) which is 

\begin{equation}
S=\int d^{4}x\sqrt{-g}R+\mathcal{L}_{m}.\label{eq: EH}\end{equation}

The action (\ref{eq: high order 1}) has been analysed in \cite{key-33,key-34,key-35}
in cosmological contexts. 

As we will interact with gravitational waves, i.e. the linearized
theory in vacuum, $\mathcal{L}_{m}=0$ will be put and the pure curvature
action \begin{equation}
S=\int d^{4}x\sqrt{-g}f(R)\label{eq: high order 12}\end{equation}
 will be considered.

\section{Field equations and linearized theory}

Following \cite{key-32} (note that in this paper we work with $8\pi G=1$,
$c=1$ and $\hbar=1$), the variational principle

\begin{equation}
\delta\int d^{4}x\sqrt{-g}f(R)=0\label{eq: high order 2}\end{equation}

in a local Lorentz frame can be used, obtaining:

\begin{equation}
f'(R)R_{\mu\nu}-\frac{1}{2}f(R)g_{\mu\nu}-f'(R)_{;\mu;\nu}+g_{\mu\nu}\square f'(R)=0\label{eq: einstein-general}\end{equation}

which are the modified Einstein field equations. $f'(R)$ is the derivative
of $f$ in respect to the Ricci scalar. Writing down, explicitly,
the Einstein tensor, eqs. (\ref{eq: einstein-general}) become\begin{equation}
G_{\mu\nu}=\frac{1}{f'(R)}\{\frac{1}{2}g_{\mu\nu}[f(R)-f'(R)R]+f'(R)_{;\mu;\nu}-g_{\mu\nu}\square f'(R)\}.\label{eq: einstein 2}\end{equation}

The trace of the field equations (\ref{eq: einstein 2}) gives 

\begin{equation}
3\square f'(R)+Rf'(R)-2f(R)=0,\label{eq: KG}\end{equation}

and, with the identifications \cite{key-40}

\begin{equation}
\begin{array}{ccccc}
\Phi\rightarrow f'(R) &  & \textrm{and } &  & \frac{dV}{d\Phi}\rightarrow\frac{2f(R)-Rf'(R)}{3}\end{array}\label{eq: identifica}\end{equation}

a Klein - Gordon equation for the effective $\Phi$ scalar field is
obtained:

\cite{key-23}\begin{equation}
\square\Phi=\frac{dV}{d\Phi}.\label{eq: KG2}\end{equation}

To study gravitational waves, the linearized theory has to be analysed,
with a little perturbation of the background, which is assumed given
by a near Minkowskian background, i.e. a Minkowskian background plus
$\Phi=\Phi_{0}$ (the Ricci scalar is assumed constant in the background)
\cite{key-9,key-19}. $\Phi_{0}$ is also assumed to be a steady minimum
for the effective potential $V$, that called $V_{0}$. This assumption
is vital for the further calculations \cite{key-41} and its physical
justification arises from the fact that the effective $\Phi$ scalar
field is a function of the Ricci curvature and here the linearized
theory is developed, i.e. only weak perturbations near a \emph{fixed}
curvature are considered. Thus, such a minimum has to be \emph{steady}.
This is an analysis \textit{totally equivalent} to the case of the
linearization process for scalar-tensor gravity, see for example \cite{key-16}.
In Section 4 of \cite{key-16} the authors claim that \textit{{}``we
linearize the equations near the background ($\eta_{\mu\nu}$, $\varphi_{0}$
), where $\varphi_{0}$ is a minimum of $V$''}. The difference with
the present analysis is that in \cite{key-16} the scalar field which
is a minimum of the potential arises directly from the Brans-Dicke
theory \cite{key-23}, while in the present analysis an effective
scalar field and an effective potential arise directly from spacetime
curvature (the effective scalar field is the prime derivative $f'(R)$,
see eq. (\ref{eq: identifica})).

Thus, the potential presents a square (i.e. parabolic) trend, in function
of the effective scalar field, near the minimum \cite{key-41}, i.e.

\begin{equation}
V\simeq V_{0}+\frac{1}{2}\alpha\delta\Phi^{2}\Rightarrow\frac{dV}{d\Phi}\simeq m^{2}\delta\Phi,\label{eq: minimo}\end{equation}

and the constant $m$ has mass dimension. 

Putting

\begin{equation}
\begin{array}{c}
g_{\mu\nu}=\eta_{\mu\nu}+h_{\mu\nu}\\
\\\Phi=\Phi_{0}+\delta\Phi.\end{array}\label{eq: linearizza}\end{equation}

to first order in $h_{\mu\nu}$ and $\delta\Phi$, calling $\widetilde{R}_{\mu\nu\rho\sigma}$
, $\widetilde{R}_{\mu\nu}$ and $\widetilde{R}$ the linearized quantities
which correspond to $R_{\mu\nu\rho\sigma}$ , $R_{\mu\nu}$ and $R$,
the linearized field equations are obtained \cite{key-12,key-19,key-31}:

\begin{equation}
\begin{array}{c}
\widetilde{R}_{\mu\nu}-\frac{\widetilde{R}}{2}\eta_{\mu\nu}=(\partial_{\mu}\partial_{\nu}h_{f}-\eta_{\mu\nu}\square h_{f})\\
\\{}\square h_{f}=m^{2}h_{f},\end{array}\label{eq: linearizzate1}\end{equation}

where 

\begin{equation}
h_{f}\equiv\frac{\delta\Phi}{\Phi_{0}}.\label{eq: definizione}\end{equation}

Then, from the second of eqs. (\ref{eq: linearizzate1}), the mass
can be defined like

\begin{equation}
m\equiv\sqrt{\frac{\square h_{f}}{h_{f}}}=\sqrt{\frac{\square\delta\Phi}{\delta\Phi}}.\label{eq: massa}\end{equation}

Thus, as the mass is generated by variation of a function of the Ricci
scalar, in a certain sense, it is generated by variation of spacetime
curvature. In this way, the theory is totally generalized for an arbitrary
$f$ function of the Ricci scalar, improving the results in \cite{key-9,key-19,key-40},
where only particular theories have been discussed.. 

$\widetilde{R}_{\mu\nu\rho\sigma}$ and eqs. (\ref{eq: linearizzate1})
are invariants for gauge transformations \cite{key-9,key-12,key-19,key-40}

\begin{equation}
\begin{array}{c}
h_{\mu\nu}\rightarrow h'_{\mu\nu}=h_{\mu\nu}-\partial_{(\mu}\epsilon_{\nu)}\\
\\\delta\Phi\rightarrow\delta\Phi'=\delta\Phi;\end{array}\label{eq: gauge}\end{equation}

then 

\begin{equation}
\bar{h}_{\mu\nu}\equiv h_{\mu\nu}-\frac{h}{2}\eta_{\mu\nu}+\eta_{\mu\nu}h_{f}\label{eq: ridefiniz}\end{equation}

can be defined, and, considering the transform for the parameter $\epsilon^{\mu}$

\begin{equation}
\square\epsilon_{\nu}=\partial^{\mu}\bar{h}_{\mu\nu},\label{eq:lorentziana}\end{equation}
 a gauge parallel to the Lorenz one of electromagnetic waves can be
chosen:

\begin{equation}
\partial^{\mu}\bar{h}_{\mu\nu}=0.\label{eq: cond lorentz}\end{equation}

In this way,the field equations read like

\begin{equation}
\square\bar{h}_{\mu\nu}=0\label{eq: onda T}\end{equation}

\begin{equation}
\square h_{f}=m^{2}h_{f}\label{eq: onda S}\end{equation}

Solutions of eqs. (\ref{eq: onda T}) and (\ref{eq: onda S}) are
plan waves \cite{key-12,key-19}:

\begin{equation}
\bar{h}_{\mu\nu}=A_{\mu\nu}(\overrightarrow{p})\exp(ip^{\alpha}x_{\alpha})+c.c.\label{eq: sol T}\end{equation}

\begin{equation}
h_{f}=a(\overrightarrow{p})\exp(iq^{\alpha}x_{\alpha})+c.c.\label{eq: sol S}\end{equation}

where

\begin{equation}
\begin{array}{ccc}
k^{\alpha}\equiv(\omega,\overrightarrow{p}) &  & \omega=p\equiv|\overrightarrow{p}|\\
\\q^{\alpha}\equiv(\omega_{m},\overrightarrow{p}) &  & \omega_{m}=\sqrt{m^{2}+p^{2}}.\end{array}\label{eq: k e q}\end{equation}

In eqs. (\ref{eq: onda T}) and (\ref{eq: sol T}) the equation and
the solution for the standard waves of General Relativity \cite{key-31,key-32}
have been obtained, while eqs. (\ref{eq: onda S}) and (\ref{eq: sol S})
are respectively the equation and the solution for the massive mode
(see also \cite{key-9,key-12,key-19,key-40}).

The fact that the dispersion law for the modes of the massive field
$h_{f}$ is not linear has to be emphasized. The velocity of every
{}``ordinary'' (i.e. which arises from General Relativity) mode
$\bar{h}_{\mu\nu}$ is the light speed $c$, but the dispersion law
(the second of eq. (\ref{eq: k e q})) for the modes of $h_{f}$ is
that of a massive field which can be discussed like a wave-packet
\cite{key-9,key-12,key-19,key-40}. Also, the group-velocity of a
wave-packet of $h_{f}$ centred in $\overrightarrow{p}$ is 

\begin{equation}
\overrightarrow{v_{G}}=\frac{\overrightarrow{p}}{\omega},\label{eq: velocita' di gruppo}\end{equation}

which is exactly the velocity of a massive particle with mass $m$
and momentum $\overrightarrow{p}$.

From the second of eqs. (\ref{eq: k e q}) and eq. (\ref{eq: velocita' di gruppo})
it is simple to obtain:

\begin{equation}
v_{G}=\frac{\sqrt{\omega^{2}-m^{2}}}{\omega}.\label{eq: velocita' di gruppo 2}\end{equation}

Then, as one wants a constant speed of the wave-packet, it has to
be \cite{key-9,key-12,key-19}

\begin{equation}
m=\sqrt{(1-v_{G}^{2})}\omega.\label{eq: relazione massa-frequenza}\end{equation}

Now, the analysis can remain in the Lorenz gauge with transformations
of the type $\square\epsilon_{\nu}=0$; this gauge gives a condition
of transverse effect for the ordinary part of the field: $k^{\mu}A_{\mu\nu}=0$,
but does not give the transverse effect for the total field $h_{\mu\nu}$.
From eq. (\ref{eq: ridefiniz}) it is

\begin{equation}
h_{\mu\nu}=\bar{h}_{\mu\nu}-\frac{\bar{h}}{2}\eta_{\mu\nu}+\eta_{\mu\nu}h_{f}.\label{eq: ridefiniz 2}\end{equation}

At this point, if being in the massless case \cite{key-9,key-12,key-19},
one puts

\begin{equation}
\begin{array}{c}
\square\epsilon^{\mu}=0\\
\\\partial_{\mu}\epsilon^{\mu}=-\frac{\bar{h}}{2}+h_{f},\end{array}\label{eq: gauge2}\end{equation}

which gives the total transverse effect of the field. But in the massive
case this is impossible. In fact, by applying the Dalembertian operator
to the second of eqs. (\ref{eq: gauge2}) and by using the field equations
(\ref{eq: onda T}) and (\ref{eq: onda S}) it is

\begin{equation}
\square\epsilon^{\mu}=m^{2}h_{f},\label{eq: contrasto}\end{equation}

which is in contrast with the first of eqs. (\ref{eq: gauge2}). In
the same way, it is possible to show that it does not exist any linear
relation between the tensor field $\bar{h}_{\mu\nu}$ and the massive
field $h_{f}$. Thus, a gauge in which $h_{\mu\nu}$ is purely spatial
cannot be chosen (i.e. $h_{\mu0}=0$ cannot be put, see eq. (\ref{eq: ridefiniz 2}))
. But the traceless condition to the field $\bar{h}_{\mu\nu}$ can
be put :

\begin{equation}
\begin{array}{c}
\square\epsilon^{\mu}=0\\
\\\partial_{\mu}\epsilon^{\mu}=-\frac{\bar{h}}{2}.\end{array}\label{eq: gauge traceless}\end{equation}

These equations imply

\begin{equation}
\partial^{\mu}\bar{h}_{\mu\nu}=0.\label{eq: vincolo}\end{equation}

To save the conditions $\partial_{\mu}\bar{h}^{\mu\nu}$ and $\bar{h}=0$
transformations like

\begin{equation}
\begin{array}{c}
\square\epsilon^{\mu}=0\\
\\\partial_{\mu}\epsilon^{\mu}=0\end{array}\label{eq: gauge 3}\end{equation}

can be used and, taking $\overrightarrow{p}$ in the $z$ direction,
a gauge in which only $A_{11}$, $A_{22}$, and $A_{12}=A_{21}$ are
different to zero can be chosen. The condition $\bar{h}=0$ gives
$A_{11}=-A_{22}$. Now, putting these equations in eq. (\ref{eq: ridefiniz 2}),
it is

\begin{equation}
h_{\mu\nu}(t,z)=A^{+}(t-z)e_{\mu\nu}^{(+)}+A^{\times}(t-z)e_{\mu\nu}^{(\times)}+h_{f}(t-v_{G}z)\eta_{\mu\nu}.\label{eq: perturbazione totale}\end{equation}

The term $A^{+}(t-z)e_{\mu\nu}^{(+)}+A^{\times}(t-z)e_{\mu\nu}^{(\times)}$
describes the two standard polarizations of gravitational waves which
arise from General Relativity, while the term $h_{f}(t-v_{G}z)\eta_{\mu\nu}$
is the massive field arising from the high order theory. In other
words, the function $f'$ of the Ricci scalar generates a third massive
polarization for gravitational waves which is not present in standard
General Relativity.

\section{The primordial production of the third polarization}

Now, let us consider the primordial physical process, which gave rise
to a characteristic spectrum $\Omega_{gw}$ for relic GWs. Such physical
process has been analysed in different ways: respectively in refs.
\cite{key-25,key-26} and \cite{key-27,key-28} but only for the components
of eq. (\ref{eq: perturbazione totale}) which arises from General
Relativity. In \cite{key-29} the process has been extended to scalar-tensor
gravity. Actually, the process can be further improved showing the
primordial production of the third polarization of eq. (\ref{eq: perturbazione totale}).

Before starting with the analysis, let us recall that, considering
a stochastic background of GWs, it can be characterized by a dimensionless
spectrum \cite{key-25,key-26,key-27,key-28,key-29}\begin{equation}
\Omega_{gw}(f)\equiv\frac{1}{\rho_{c}}\frac{d\rho_{gw}}{d\ln f},\label{eq: spettro}\end{equation}

where \begin{equation}
\rho_{c}\equiv\frac{3H_{0}^{2}}{8G}\label{eq: densita' critica}\end{equation}

is the (actual) critical density energy, $\rho_{c}$ of the Universe,
$H_{0}$ the actual value of the Hubble expansion rate and $d\rho_{gw}$
the energy density of relic GWs in the frequency range $f$ to $f+df$.

The existence of a relic stochastic background of GWs arises from
generals assumptions, i.e. from a mixing between basic principles
of classical theories of gravity and of quantum field theory. The
strong variations of the gravitational field in the early universe
amplify the zero-point quantum oscillations and produce relic GWs.
It is well known that the detection of relic GWs is the only way to
learn about the evolution of the very early universe, up to the bounds
of the Planck epoch and the initial singularity \cite{key-21,key-25,key-26,key-27,key-28,key-29}.
It is very important to stress the unavoidable and fundamental character
of this mechanism. The model derives from the inflationary scenario
for the early universe \cite{key-36,key-37}, which is tuned in a
good way with the WMAP data on the Cosmic Background Radiation (CBR)
(in particular exponential inflation and spectral index $\approx1$
\cite{key-38,key-39}). Inflationary models of the early Universe
were analysed in the early and middles 1980's (see \cite{key-36}
for a review ), starting from an idea of A. Guth \cite{key-37}. These
are cosmological models in which the Universe undergoes a brief phase
of a very rapid expansion in early times. In this context, the expansion
could be power-law or exponential in time. Inflationary models provide
solutions to the horizon and flatness problems and contain a mechanism
which creates perturbations in all fields. Important for our goals
is that this mechanism also provides a distinctive spectrum of relic
GWs. The GWs perturbations arise from the uncertainty principle and
the spectrum of relic GWs is generated from the adiabatically-amplified
zero-point fluctuations \cite{key-21,key-25,key-26,key-27,key-28,key-29}. 

Now, the calculation for a simple inflationary model will be shown
for the third polarization of eq. (\ref{eq: perturbazione totale}),
following the works of Allen \cite{key-25,key-26} that performed
the calculation in the case of standard General Relativity and Corda,
Capozziello and De Laurentis \cite{key-29,key-30} that extended the
process to scalar GWs. In a recent paper \cite{key-21}, such a process
has been applied to the $f(R)$ theories arising from the action (\ref{eq: high order 1}).
But, in \cite{key-21} a different point of view has been considered.
In that case, using a conform analysis, the authors discussed such
a process in respect to the two standard polarizations which arises
from standard General Relativity. In the following the analysis is
focused to the third massive polarization. Thus, in a certain sense,
one can say that the present analysis is an integration of the analysis
in \cite{key-21}.

It will be assumed that the universe is described by a simple cosmology
in two stages, an inflationary De Sitter phase and a radiation dominated
phase \cite{key-21,key-25,key-26,key-27,key-28,key-29}. The line
element of the spacetime is given by

\begin{equation}
ds^{2}=a^{2}(\eta)[-d\eta^{2}+d\overrightarrow{x}^{2}+h_{\mu\nu}(\eta,\overrightarrow{x})dx^{\mu}dx^{\nu}].\label{eq: metrica}\end{equation}

(\ref{eq: metrica}) has to be a solution of the general field equations
(6). In fact, even if such a form is allowed, it could have absolutely
different behavior (see \cite{key-44,key-45,key-46}). For instance
one needs to show that inflation is present in the proposed model.
But, in the case of f(R) theories, which are the ones that we are
treating here, both of the two conditions are, in general, satisfied.
In fact, we recall that the original inflation was proposed by Starobinsky
in the classical papers \cite{key-48,key-49} by using the simplest
f(R) theory, i.e. the $R^{2}$ one. On the other hand, the potential
presence and the importance of standard De Sitter inflation in the
general framework of the primordial production of relic gravitational
waves has been recently shown in \cite{key-50} considering a different
point of view. In that case, using a conform analysis, the authors
discussed such a process in respect to the two standard polarizations
which arises from standard General Relativity. In the following, the
analysis is focused to the third massive polarization. Thus, in a
certain sense, the present analysis is an integration of the analysis
in \cite{key-50}. Remaining in the tapestry of f(R) theories, there
are lots of examples in the literature where the line element (\ref{eq: metrica}),
which is the perturbed form of the standard conformally flat Robertson-Walker
one, results a solution of the general field equations (6). One can
see for example the recent reviews \cite{key-51,key-52,key-53}.

In the line element (\ref{eq: metrica}), by considering only the
third polarization, the metric perturbation (\ref{eq: perturbazione totale})
reduces to 

\begin{equation}
h_{\mu\nu}=h_{f}I_{\mu\nu},\label{eq: perturbazione scalare}\end{equation}

where \begin{equation}
I_{\mu\nu}\equiv\begin{array}{cccc}
1 & 0 & 0 & 0\\
0 & 1 & 0 & 0\\
0 & 0 & 1 & 0\\
0 & 0 & 0 & 1.\end{array}\label{eq: identica}\end{equation}

In the De Sitter phase ($\eta<\eta_{1}$) the equation of state is
$P=-\rho=const$, the scale factor is $a(\eta)=\eta_{1}^{2}\eta_{0}^{-1}(2\eta_{1}-\eta)^{-1}$
and the Hubble constant is given by $H(\eta)=H_{ds}=c\eta_{0}/\eta_{1}^{2}$.

In the radiation dominated phase $(\eta>\eta_{1})$ the equation of
state is $P=\rho/3$, the scale factor is $a(\eta)=\eta/\eta_{0}$
and the Hubble constant is given by $H(\eta)=c\eta_{0}/\eta^{2}$~\cite{key-21,key-25,key-26,key-29,key-30}.

Expressing the scale factor in terms of comoving time defined by

\begin{equation}
cdt=a(t)d\eta\label{eq: tempo conforme}\end{equation}

one gets

\begin{equation}
a(t)\propto\exp(H_{ds}t)\label{eq: inflazione}\end{equation}

during the De Sitter phase and

\begin{equation}
a(t)\propto\sqrt{t}\label{eq: dominio radiazione}\end{equation}

during the radiation dominated phase. The horizon and flatness problems
are solved if \cite{key-36,key-37}

\begin{center}
$\frac{a(\eta_{0})}{a(\eta_{1})}>10^{27}$
\par\end{center}

The third polarization generates weak perturbations $h_{\mu\nu}(\eta,\overrightarrow{x})$
of the metric (\ref{eq: perturbazione scalare}) that can be written,
in terms of the conformal time $\eta$, in the form 

\begin{equation}
h_{\mu\nu}=I_{\mu\nu}(\hat{k})X(\eta)\exp(\overrightarrow{k}\cdot\overrightarrow{x}),\label{eq: relic gravity-waves}\end{equation}

where $\overrightarrow{k}$ is a constant wavevector and

\begin{equation}
h_{f}(\eta,\overrightarrow{k},\overrightarrow{x})=X(\eta)\exp(\overrightarrow{k}\cdot\overrightarrow{x}).\label{eq: phi}\end{equation}
By putting $Y(\eta)=a(\eta)X(\eta)$  one performs the standard linearized
calculation in which the connections (i.e. the Cristoffel coefficients),
the Riemann tensor, the Ricci tensor and the Ricci scalar curvature
are computed. Then, from the Friedman linearized equations, the function
$Y(\eta)$ satisfies the equation

\begin{equation}
Y''+(|\overrightarrow{k}|^{2}-\frac{a''}{a})Y=0\label{eq: Klein-Gordon}\end{equation}

where $'$ denotes derivative with respect to the conformal time.
Clearly, this is the equation for a parametrically disturbed oscillator.

The solutions of eq. (\ref{eq: Klein-Gordon}) give the solutions
for the function $X(\eta)$, that can be expressed in terms of elementary
functions simple cases of half integer Bessel or Hankel functions
\cite{key-21,key-25,key-26,key-29,key-30} in both of the inflationary
and radiation dominated eras:

For $\eta<\eta_{1}$ \begin{equation}
X(\eta)=\frac{a(\eta_{1})}{a(\eta)}[1+H_{ds}\omega^{-1}]\exp-ik(\eta-\eta_{1}),\label{eq: ampiezza inflaz.}\end{equation}

for $\eta>\eta_{1}$ \begin{equation}
X(\eta)=\frac{a(\eta_{1})}{a(\eta)}[\alpha\exp-ik(\eta-\eta_{1})+\beta\exp ik(\eta-\eta_{1}),\label{eq: ampiezza rad.}\end{equation}

where $\omega=ck/a$ is the angular frequency of the wave (that is
function of the time because of the constance of $k=|\overrightarrow{k}|$),
$\alpha$ and $\beta$ are time-independent constants which can be
obtained demanding that both $X$ and $dX/d\eta$ are continuous at
the boundary $\eta=\eta_{1}$ between the inflationary and the radiation
dominated eras of the cosmological expansion. With this constrain
it is

\begin{equation}
\alpha=1+i\frac{\sqrt{H_{ds}H_{0}}}{\omega}-\frac{H_{ds}H_{0}}{2\omega^{2}}\label{eq: alfa}\end{equation}

\begin{equation}
\beta=\frac{H_{ds}H_{0}}{2\omega^{2}}\label{eq: beta}\end{equation}

In eqs. (\ref{eq: alfa}), (\ref{eq: beta}) $\omega=ck/a(\eta_{0})$
is the angular frequency that would be observed today. Calculations
like this are referred in the literature as Bogoliubov coefficient
methods \cite{key-21,key-25,key-26,key-29,key-30}. 

As inflation damps out any classical or macroscopic perturbations,
the minimum allowed level of fluctuations is that required by the
uncertainty principle. The solution (\ref{eq: ampiezza inflaz.})
corresponds precisely to this De Sitter vacuum state \cite{key-21,key-25,key-26,key-29,key-30}.
Then, if the period of inflation was long enough, the observable properties
of the Universe today should be the same properties of a Universe
started in the De Sitter vacuum state.

In the radiation dominated phase the coefficients of $\alpha$ are
the eigenmodes which describe particles while the coefficients of
$\beta$ are the eigenmodes which describe antiparticles. Thus, the
number of created particles of angular frequency $\omega$ in this
phase is 

\begin{equation}
N_{\omega}=|\beta_{\omega}|^{2}=(\frac{H_{ds}H_{0}}{2\omega^{2}})^{2}.\label{eq: numero quanti}\end{equation}

Now, one can write an expression for the energy spectrum of the relic
gravitational waves background in the frequency interval $(\omega,\omega+d\omega)$
as

\begin{equation}
d\rho_{gw}=2\hbar\omega(\frac{\omega^{2}d\omega}{2\pi^{2}c^{3}})N_{\omega}=\frac{\hbar H_{ds}^{2}H_{0}^{2}}{4\pi^{2}c^{3}}\frac{d\omega}{\omega}=\frac{\hbar H_{ds}^{2}H_{0}^{2}}{4\pi^{2}c^{3}}\frac{df}{f}.\label{eq: de energia}\end{equation}

Eq. (\ref{eq: de energia}) can be rewritten in terms of the present
day and the De Sitter energy-density of the Universe. The Hubble expansion
rates is

\begin{center}
$H_{0}^{2}=\frac{8\pi G\rho_{c}}{3}$, $H_{ds}^{2}=\frac{8\pi G\rho_{ds}}{3}$.
\par\end{center}

Then, introducing the Planck density \begin{equation}
\rho_{Planck}\equiv\frac{c^{7}}{\hbar G^{2}}\label{eq: Shoooortyyyyyy}\end{equation}
 the spectrum is 

\begin{equation}
\Omega_{gw}(f)=\frac{1}{\rho_{c}}\frac{d\rho_{sgw}}{d\ln f}=\frac{f}{\rho_{c}}\frac{d\rho_{gw}}{df}=\frac{16}{9}\frac{\rho_{ds}}{\rho_{Planck}}.\label{eq: spettro gravitoni}\end{equation}

Some comments are needed. The computation works for a very simplified
model that does not include the matter dominated era. Including this
era, the redshift has to be considered. An enlightening computation
parallel to the one in \cite{key-26} gives

\begin{equation}
\Omega_{gw}(f)=\frac{16}{9}\frac{\rho_{ds}}{\rho_{Planck}}(1+z_{eq})^{-1},\label{eq: spettro gravitoni redshiftato}\end{equation}

for the waves which at the time in which the Universe was becoming
matter dominated had a frequency higher than $H_{eq}$, the Hubble
constant at that time. This corresponds to frequencies $f>(1+z_{eq})^{1/2}H_{0}$,
where $z_{eq}$ is the redshift of the Universe when the matter and
radiation energy density were equal. The redshift correction in equation
(\ref{eq: spettro gravitoni redshiftato}) is needed because the Hubble
parameter, which is governed by Friedman equations, should be different
from the observed one $H_{0}$ for a Universe without matter dominated
era.

At lower frequencies the spectrum is \cite{key-21,key-25,key-26,key-29,key-30}

\begin{equation}
\Omega_{gw}(f)\propto f^{-2}.\label{eq: spettro basse frequenze}\end{equation}

Moreover, the results (\ref{eq: spettro gravitoni}) and (\ref{eq: spettro gravitoni redshiftato}),
which are not frequency dependent, cannot be applied to all the frequencies.
For waves with frequencies less than $H_{0}$ today, the energy density
cannot be defined, because the wavelength becomes longer than the
Hubble radius. In the same way, at high frequencies there is a maximum
frequency above which the spectrum drops to zero rapidly. In the above
computation it has been implicitly assumed that the phase transition
from the inflationary to the radiation dominated epoch is instantaneous.
In the real Universe this phase transition occurs over some finite
time $\Delta\tau$, and above a frequency

\begin{equation}
f_{max}=\frac{a(t_{1})}{a(t_{0})}\frac{1}{\Delta\tau},\label{eq: freq. max}\end{equation}

which is the redshifted rate of the transition, $\Omega_{gw}$ drops
rapidly. These two cutoffs, at low and high frequencies, to the spectrum
force the total energy density of the relic gravitational waves to
be finite. For GUT energy-scale inflation it is \cite{key-21,key-25,key-26,key-29,key-30}. 

\begin{equation}
\frac{\rho_{ds}}{\rho_{Planck}}\approx10^{-12}.\label{eq: rapporto densita' primordiali}\end{equation}

\section{Tuning with WMAP data}

It is well known that WMAP observations put strongly severe restrictions
on the spectrum of relic gravitational waves. In fig. 1 the spectrum
$\Omega_{gw}$is mapped following \cite{key-20}: the amplitude is
chosen (determined by the ratio $\frac{\rho_{ds}}{\rho_{Planck}}$)
to be \textit{as large as possible, consistent with the WMAP constraints}
o\textit{n tensor perturbations.} Nevertheless, because the spectrum
falls off $\propto f^{-2}$ at low frequencies, this means that today,
at LIGO-Virgo and LISA frequencies (indicate by the lines in fig.
1) \cite{key-20}, it is

\begin{center}
\begin{equation}
\Omega_{gw}(f)h_{100}^{2}<9*10^{-13}.\label{eq: limite spettro WMAP}\end{equation}

\par\end{center}

Let us calculate the correspondent strain at $\approx100Hz$, where
interferometers like Virgo and LIGO have a maximum in sensitivity.
The well known equation for the characteristic amplitude, adapted
for the third component of GWs can be used \cite{key-20}:

\begin{equation}
h_{fc}(f)\simeq1.26*10^{-18}(\frac{1Hz}{f})\sqrt{h_{100}^{2}\Omega_{gw}(f)},\label{eq: legame ampiezza-spettro}\end{equation}
obtaining \cite{key-20}

\begin{equation}
h_{fc}(100Hz)<1.7*10^{-26}.\label{eq: limite per lo strain}\end{equation}

Then, as we expect a sensitivity order of $10^{-22}$ for interferometers
at $\approx100Hz$, four order of magnitude have to be gained. Let
us analyse smaller frequencies too. The sensitivity of the Virgo interferometer
is of the order of $10^{-21}$ at $\approx10Hz$ and in that case
it is \cite{key-20}

\begin{equation}
h_{fc}(10Hz)<1.7*10^{-25}.\label{eq: limite per lo strain2}\end{equation}

The sensitivity of the LISA interferometer will be of the order of
$10^{-22}$ at $10^{-3}\approx Hz$ and in that case it is \cite{key-20}

\begin{equation}
h_{fc}(100Hz)<1.7*10^{-21}.\label{eq: limite per lo strain3}\end{equation}

Then, a stochastic background of relic gravitational waves could be,
in principle, detected by the LISA interferometer.

We emphasize the sentence \emph{in principle}. Actually, one has to
take into account the Galactic confusion background which will dominate
over the instrumental noise. Thus, it is still questionable whether
the relic gravitons produced in GR could be detected by LISA \cite{key-41}.

On the other hand, the assumption that all the tensor perturbations
in the Universe are due to a stochastic background of GWs is quit
strong, but the results (\ref{eq: limite spettro WMAP}), (\ref{eq: limite per lo strain}),
(\ref{eq: limite per lo strain2}) and (\ref{eq: limite per lo strain3})
can be considered like upper bounds.

Figure 1, which is adapted from ref. \cite{key-20}, shows that the
spectrum of relic SGWs in inflationary models is flat over a wide
range of frequencies. The horizontal axis is $\log_{10}$ of frequency,
in Hz. The vertical axis is $\log_{10}\Omega_{gsw}$. The inflationary
spectrum rises quickly at low frequencies (wave which re-entered in
the Hubble sphere after the Universe became matter dominated) and
falls off above the (appropriately redshifted) frequency scale $f_{max}$
associated with the fastest characteristic time of the phase transition
at the end of inflation. The amplitude of the flat region depends
only on the energy density during the inflationary stage; we have
chosen the largest amplitude consistent with the WMAP constrains on
tensor perturbations. This means that at LIGO and LISA frequencies,
$\Omega_{gw}(f)h_{100}^{2}<9*10^{-13}.$

\begin{figure}

\includegraphics{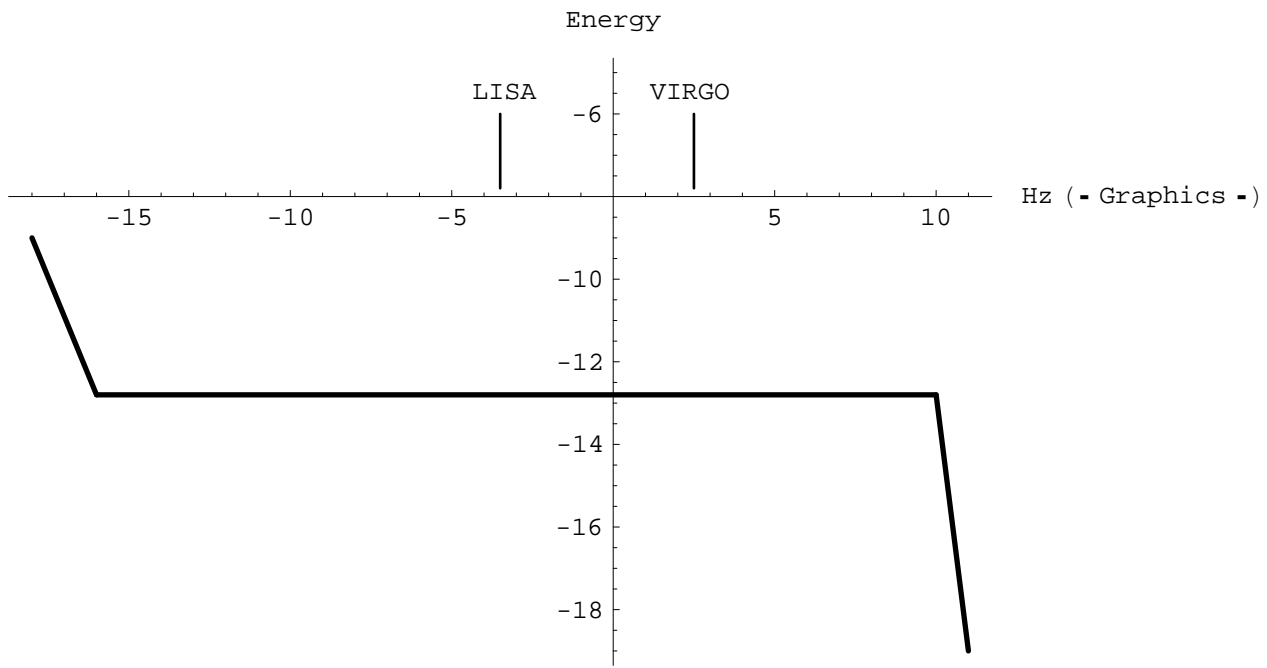}\caption{The spectrum of relic SGWs in inflationary models is flat over a wide
range of frequencies. The horizontal axis is $\log_{10}$ of frequency,
in Hz. The vertical axis is $\log_{10}\Omega_{gsw}$.}

\end{figure}

\section{Potential detection with interferometers}

Before starting the discussion of the potential interferometric detection
of the massive GWs polarization, there is another point which has
to be clarified. The attentive reader \cite{key-41} asks what would
happen to Hulse-Taylor pulsar \cite{key-54}. As it is clear from
eq. (\ref{eq: perturbazione totale}), by fixing the massless component
of the metric to GR value, the third term should carry the additional
energy which should affect evolution of the binary system \cite{key-41}.
This problem has been discussed by Shibata, Nakao and Nakamura in
\cite{key-15}. In such a work, it has been shown that the energy
associated to the monopole mode of the scalar particle is quite lower
in respect to the energy carried by ordinary quadrupole modes arising
from the linearization of standard General Relativity. Thus, in these
conditions, the evolution of the binary system should be not affected
by the third polarization if one remains into observing error bars.

On the other and, even if the ratio of differential mode to common
mode is large in the case of the binary pulsar, it could be, in principle,
similar and potentially small in cosmology. In fact, in the case of
$f(R)$ theories such a mode arises directly from a function of the
Ricci scalar, see the first identification (\ref{eq: identifica}),
i.e. \textit{it arises from spacetime curvature}, and it is well known
that, being the production of relic GWs near the bounds of the Planck
epoch and the initial singularity, in this early Era both of spacetime
curvature and its variations were very high. Thus, in this cosmological
case, the amplitude of the common mode could be even dominant in respect
to differential modes. 

After these clarifications, let us start the discussion regarding
the potential detection. 

Even if eqs. (\ref{eq: limite per lo strain}) and (\ref{eq: limite per lo strain2})
show that the detection of relic GWs is quite difficult, people hope
in better sensitivity of advanced projects. In this case, it is interesting
to discuss the interaction between interferometers and massive GWs.

Considering only the third polarization $h_{f}(t-v_{G}z)\eta_{\mu\nu}$
the line element associated to eq. (\ref{eq: perturbazione totale})
becomes the conformally flat one

\begin{equation}
ds^{2}=[1+h_{f}(t,z)](-dt^{2}+dz^{2}+dx^{2}+dy^{2}).\label{eq: metrica puramente scalare}\end{equation}
As the analysis on the motion of test masses is performed in a laboratory
environment on Earth, the coordinate system in which the space-time
is locally flat is typically used and the distance between any two
points is given simply by the difference in their coordinates in the
sense of Newtonian physics \cite{key-12,key-13,key-14,key-16,key-31}.
This frame is the proper reference frame of a local observer, located
for example in the position of the beam splitter of an interferometer.
In this frame GWs manifest themself by exerting tidal forces on the
masses (the mirror and the beam-splitter in the case of an interferometer).
A detailed analysis of the frame of the local observer is given in
ref. \cite{key-31}, sect. 13.6. Here only the more important features
of this coordinate system are recalled:

the time coordinate $x_{0}$ is the proper time of the observer O;

spatial axes are centred in O;

in the special case of zero acceleration and zero rotation the spatial
coordinates $x_{j}$ are the proper distances along the axes and the
frame of the local observer reduces to a local Lorentz frame: in this
case the line element reads \cite{key-31}

\begin{equation}
ds^{2}=-(dx^{0})^{2}+\delta_{ij}dx^{i}dx^{j}+O(|x^{j}|^{2})dx^{\alpha}dx^{\beta}.\label{eq: metrica local lorentz}\end{equation}

The effect of the gravitational wave on test masses is described by
the equation

\begin{equation}
\ddot{x^{i}}=-\widetilde{R}_{0k0}^{i}x^{k},\label{eq: deviazione geodetiche}\end{equation}
which is the equation for geodesic deviation in this frame.

Thus, to study the effect of the massive gravitational wave on test
masses, $\widetilde{R}_{0k0}^{i}$ have to be computed in the proper
reference frame of the local observer. But, because the linearized
Riemann tensor $\widetilde{R}_{\mu\nu\rho\sigma}$ is invariant under
gauge transformations \cite{key-9,key-12,key-13,key-14,key-31}, it
can be directly computed from eq. (\ref{eq: perturbazione scalare}). 

From \cite{key-31} it is:

\begin{equation}
\widetilde{R}_{\mu\nu\rho\sigma}=\frac{1}{2}\{\partial_{\mu}\partial_{\beta}h_{\alpha\nu}+\partial_{\nu}\partial_{\alpha}h_{\mu\beta}-\partial_{\alpha}\partial_{\beta}h_{\mu\nu}-\partial_{\mu}\partial_{\nu}h_{\alpha\beta}\},\label{eq: riemann lineare}\end{equation}

that, in the case eq. (\ref{eq: perturbazione scalare}), begins

\begin{equation}
\widetilde{R}_{0\gamma0}^{\alpha}=\frac{1}{2}\{\partial^{\alpha}\partial_{0}h_{f}\eta_{0\gamma}+\partial_{0}\partial_{\gamma}h_{f}\delta_{0}^{\alpha}-\partial^{\alpha}\partial_{\gamma}h_{f}\eta_{00}-\partial_{0}\partial_{0}h_{f}\delta_{\gamma}^{\alpha}\};\label{eq: riemann lin scalare}\end{equation}
the different elements are (only the non zero ones will be written):

\begin{equation}
\partial^{\alpha}\partial_{0}h_{f}\eta_{0\gamma}=\left\{ \begin{array}{ccc}
\partial_{t}^{2}h_{f} & for & \alpha=\gamma=0\\
\\-\partial_{z}\partial_{t}h_{f} & for & \alpha=3;\gamma=0\end{array}\right\} \label{eq: calcoli}\end{equation}

\begin{equation}
\partial_{0}\partial_{\gamma}h_{f}\delta_{0}^{\alpha}=\left\{ \begin{array}{ccc}
\partial_{t}^{2}h_{f} & for & \alpha=\gamma=0\\
\\\partial_{t}\partial_{z}h_{f} & for & \alpha=0;\gamma=3\end{array}\right\} \label{eq: calcoli2}\end{equation}

\begin{equation}
-\partial^{\alpha}\partial_{\gamma}h_{f}\eta_{00}=\partial^{\alpha}\partial_{\gamma}h_{f}=\left\{ \begin{array}{ccc}
-\partial_{t}^{2}h_{f} & for & \alpha=\gamma=0\\
\\\partial_{z}^{2}h_{f} & for & \alpha=\gamma=3\\
\\-\partial_{t}\partial_{z}h_{f} & for & \alpha=0;\gamma=3\\
\\\partial_{z}\partial_{t}h_{f} & for & \alpha=3;\gamma=0\end{array}\right\} \label{eq: calcoli3}\end{equation}

\begin{equation}
-\partial_{0}\partial_{0}h_{f}\delta_{\gamma}^{\alpha}=\begin{array}{ccc}
-\partial_{z}^{2}h_{f} & for & \alpha=\gamma\end{array}.\label{eq: calcoli4}\end{equation}

Now, putting these results in eq. (\ref{eq: riemann lin scalare})
one obtains:

\begin{equation}
\begin{array}{c}
\widetilde{R}_{010}^{1}=-\frac{1}{2}\ddot{h}_{f}\\
\\\widetilde{R}_{010}^{2}=-\frac{1}{2}\ddot{h}_{f}\\
\\\widetilde{R}_{030}^{3}=\frac{1}{2}\square h_{f}.\end{array}\label{eq: componenti riemann}\end{equation}

But, putting the second of equations (\ref{eq: linearizzate1}) in
the third of eqs. (\ref{eq: componenti riemann}), it is

\begin{equation}
\widetilde{R}_{030}^{3}=\frac{1}{2}m^{2}h_{f},\label{eq: terza riemann}\end{equation}

which shows that the field is not transversal. 

In fact, using eq. (\ref{eq: deviazione geodetiche}) it results

\begin{equation}
\ddot{x}=\frac{1}{2}\ddot{h}_{f}x,\label{eq: accelerazione mareale lungo x}\end{equation}

\begin{equation}
\ddot{y}=\frac{1}{2}\ddot{h}_{f}y\label{eq: accelerazione mareale lungo y}\end{equation}

and 

\begin{equation}
\ddot{z}=-\frac{1}{2}m^{2}h_{f}(t,z)z.\label{eq: accelerazione mareale lungo z}\end{equation}

Then, the effect of the mass is the generation of a \textit{longitudinal}
force (in addition to the transverse one). 

For a better understanding of this longitudinal force, let us analyse
the effect on test masses in the context of the geodesic deviation.

Following \cite{key-14} one puts\begin{equation}
\widetilde{R}_{0j0}^{i}=\frac{1}{2}\left(\begin{array}{ccc}
-\partial_{t}^{2} & 0 & 0\\
0 & -\partial_{t}^{2} & 0\\
0 & 0 & m^{2}\end{array}\right)h_{f}(t,z)=-\frac{1}{2}T_{ij}\partial_{t}^{2}h_{f}+\frac{1}{2}L_{ij}m^{2}h_{f}.\label{eq: eqs}\end{equation}

Here the transverse projector with respect to the direction of propagation
of the GW $\widehat{n}$, defined by

\begin{equation}
T_{ij}=\delta_{ij}-\widehat{n}_{i}\widehat{n}_{j},\label{eq: Tij}\end{equation}

and the longitudinal projector defined by

\begin{equation}
L_{ij}=\widehat{n}_{i}\widehat{n}_{j}\label{eq: Lij}\end{equation}
have been used. In this way, the geodesic deviation equation (\ref{eq: deviazione geodetiche})
can be rewritten like

\begin{equation}
\frac{d^{2}}{dt^{2}}x_{i}=\frac{1}{2}\partial_{t}^{2}h_{f}T_{ij}x_{j}-\frac{1}{2}m^{2}h_{f}L_{ij}x_{j}.\label{eq: TL}\end{equation}

Thus, it appears clear that the effect of the mass present in the
GW generates a longitudinal force proportional to $m^{2}$ which is
in addition to the transverse one. But if $v(\omega)\rightarrow1$
in eq. (\ref{eq: velocita' di gruppo 2}) one gets $m\rightarrow0$,
and the longitudinal force vanishes. Then, it is clear that the longitudinal
mode arises from the fact that the GW does no propagate at the speed
of light.

Now, let us analyse the detectability of the third polarization computing
the pattern function of a detector to this massive component. One
has to recall that it is possible to associate to a detector a \textit{detector
tensor} that, for an interferometer with arms along the $\hat{u}$
e $\hat{v}$ directions in respect to the propagating gravitational
wave (see figure 2), is defined by \cite{key-14}\begin{equation}
D^{ij}\equiv\frac{1}{2}(\hat{v}^{i}\hat{v}^{j}-\hat{u}^{i}\hat{u}^{j}).\label{eq: definizione D}\end{equation}

\begin{figure}

\includegraphics{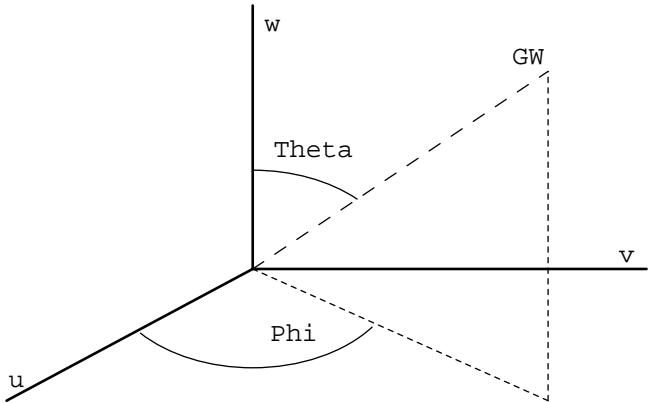}\caption{A gravitational wave propagating in the z direction}

\end{figure}
If the detector is an interferometer \cite{key-1,key-2,key-3,key-4,key-5,key-6,key-7,key-8,key-21},
the signal induced by a GW of a generic polarization, here labelled
with $s(t),$ is the phase shift, which is proportional to \cite{key-14}

\begin{equation}
s(t)\sim D^{ij}\widetilde{R}_{i0j0}\label{eq: legame onda-output}\end{equation}

and, using equations (\ref{eq: eqs}), one gets

\begin{equation}
s(t)\sim-\sin^{2}\theta\cos2\phi.\label{eq: legame onda-output 2}\end{equation}

The angular dependence (\ref{eq: legame onda-output 2}), is different
from the two well known standard ones arising from general relativity
which are, respectively $(1+\cos^{2}\theta)\cos2\phi$ for the $+$
polarization and $-\cos\theta\sin2\vartheta$ for the $\times$ polarization.
Thus, in principle, the angular dependence (\ref{eq: legame onda-output 2})
could be used to discriminate among $f(R)$ theories and general relativity,
if present or future detectors will achieve a high sensitivity. The
third angular dependence is shown in figure.

\begin{figure}

\includegraphics{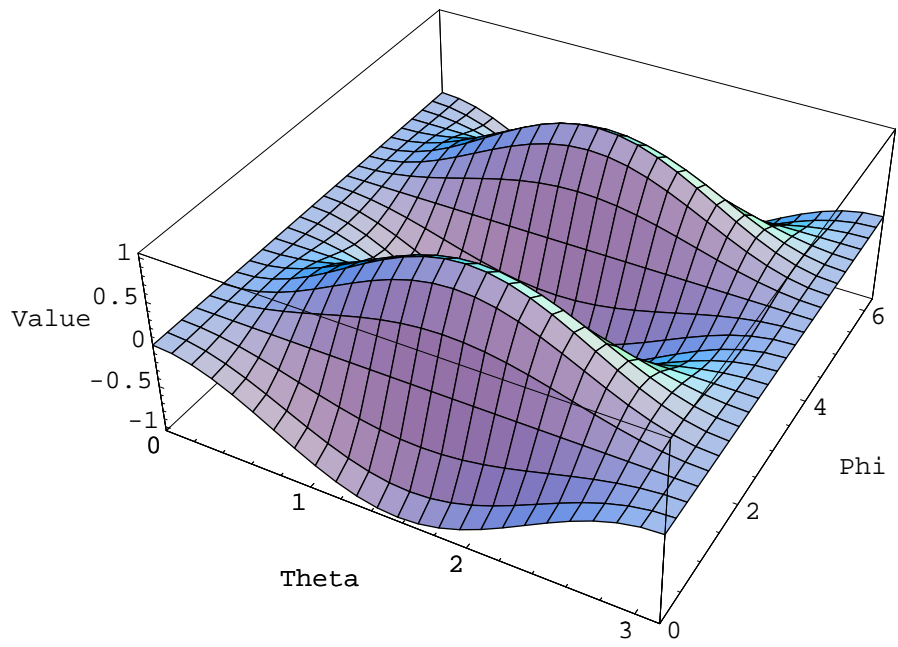}\caption{The (dimensionless) angular dependence (\ref{eq: legame onda-output 2})}

\end{figure}

For a sake of completeness, let us recall that there is one more problem
with the potential detection of the common mode. This is because the
sensitivity curve drawn for different detectors is for the differential
mode. The common mode is also registered during the GW experiment
but it is much noisier, therefore it has different sensitivity level
\cite{key-41}. Actually, this is correct for studies on potential
detection in the case of wavelength of the wave much larger than the
linear dimension of the interferometer (low frequencies approximation).
In previous analysis, we have implicitly assumed to be in such an
approximation, and in particular, eqs. (\ref{eq: definizione D}),
(\ref{eq: legame onda-output}) and (\ref{eq: legame onda-output 2})
are strictly valid only in this approximation. A recent analysis \cite{key-55}
has shown that, in the case in which the differential mode is massive,
the response function of an interferometer \textit{increases} with
increasing frequency differently from the case of massless modes where
the response function \textit{decreases} with increasing frequency.
This opens important perspectives for a potential detection of the
massive mode at high frequency. In \cite{key-55} the response function
has only been computed in the case of a massive mode propagating parallel
to one arm of the interferometer, thus, further studies in this direction
are needed. For example, it will be quite interesting to generalize
the frequency dependence of the angular pattern (\ref{eq: legame onda-output 2}).

\section{Conclusion remarks}

It has been shown that, in general, $f(R)$ theories produce a third
massive polarization of gravitational waves and the primordial production
of this polarization has been analysed adapting the adiabatically-amplified
zero-point fluctuations process at this case and generalizing previous
results in which only particular cases have been discussed. 

The presence of the mass could also have important applications in
cosmology because the fact that gravitational waves can have mass
could give a contribution to the dark matter of the Universe.

An upper bound for these relic gravitational waves, which arises from
the WMAP constrains, has been also released and at the end of the
paper and the potential detection of such massive GWs using interferometers
like Virgo and LIGO has been discussed.

\section*{Acknowledgements}

I would like to thank Francesco Rubanu for helpful advices during
my work. I have to strongly thank the referees for useful advices
and precious comments which permitted in improving the paper.

\end{document}